\documentclass{aastex}
\usepackage{emulateapj5}

\shorttitle{CorMASS} \shortauthors{Wilson et al.}

\begin{document}
\submitted{PASP, accepted 30 Oct 2000}
\title{CorMASS: A Compact and Efficient
NIR Spectrograph for Studying Low-Mass Objects\altaffilmark{1}}

\author{J.C. Wilson\altaffilmark{2},
M.F. Skrutskie\altaffilmark{3}, M.R. Colonno\altaffilmark{4},
A.T. Enos\altaffilmark{4}, J.D. Smith\altaffilmark{4},
C.P. Henderson\altaffilmark{4}, J.E. Gizis\altaffilmark{5},
D.G. Monet\altaffilmark{6} and J.R. Houck\altaffilmark{4}}

\altaffiltext{1}{Observations made at the Palomar Observatory were
made as part of a continuing collaboration between the California
Institute of Technology and Cornell University. The 60-inch
telescope at Palomar Mountain is jointly owned by the California
Institute of Technology and the Carnegie Institution of
Washington.}
\altaffiltext{2}{Space Sciences Building, Cornell
University, Ithaca, NY 14853; jcw14@cornell.edu}
\altaffiltext{3}{Department of Astronomy, University of
    Massachusetts, Amherst, MA 01003}
\altaffiltext{4}{Space Sciences Building, Cornell University,
    Ithaca, NY 14853}
\altaffiltext{5}{Infrared Processing and Analysis Center, M/S
    100-22, California Institute of Technology, Pasadena, CA 91125}
\altaffiltext{6}{U.S. Naval Observatory, P.O. Box 1149, Flagstaff,
    AZ 86002}

\begin{abstract}
CorMASS (Cornell Massachusetts Slit Spectrograph) is a compact,
low-resolution ($R\sim300$), double-pass prism cross-dispersed
near-infrared (NIR) spectrograph in operation on the Palomar
Observatory 60-inch telescope.  Its 2-dimensional spectral format
provides simultaneous coverage from $\lambda\sim 0.75 \micron$ to
$\lambda \sim 2.5 \micron$ (\textit{\'{z}JHK} bands).  A remotely
operated cold flip mirror permits its NICMOS\,3 detector to
function as a $K_s$ slit viewer to assist object placement into
the $2\arcsec \times 15\arcsec$ slit.  CorMASS was primarily
designed for the rapid spectral classification of low-mass stellar
and sub-stellar objects identified by the Two-Micron All Sky
Survey (2MASS).  CorMASS' efficiency and resolution also make it a
versatile instrument for the spectral observation and
classification of many other types of bright objects ($K<14$)
including quasars, novae, and emission line objects.
\end{abstract}


\section{Introduction}
Recent successes by deep near-infrared (NIR) surveys such as the
Two Micron All Sky Survey (2MASS; \citep{skr97}), the Deep
Near-Infrared Survey (DENIS; \citep{epc97}), and the Sloan Digital
Sky Survey (SDSS; \citep{gun95}) in identifying very low-mass
stellar and sub-stellar field candidates have made a tremendous
impact upon the study of low-mass objects in the local stellar
neighborhood.  Spectroscopic follow-up of low-mass survey
candidates has revealed objects with atmospheric features
different from the TiO and VO features seen in the atmospheres of
M-dwarfs.  Two new classifications, L with $1300 \la T_{eff} \la
2000$ \citep{ki00a} and T with $T_{eff} \la 1200-1300$
\citep{feg96,bur99,ki00a}, have been proposed to account for the
unique spectral characteristics of these objects at and below the
bottom of the main sequence \citep{mar97,kir99}.  L-dwarf spectra
are dominated by metallic hydrides and alkali atomic features as
the oxides of M-dwarfs condense into solids at these cooler
atmospheric temperatures.  T-dwarfs show the distinctive ${\rm
CH}_4$ absorption features at $1.6 \micron$ and $2.2 \micron$.  Gl
229B is the archetypical T-dwarf \citep{nak95,opp98}.

Motivated by the potential identification of a few $10^3$
candidate L and T-dwarfs in the northern sky by 2MASS
\citep{rei99}, we have built a spectrograph specifically for the
efficient confirmation and classification of low-mass object
candidates.  This paper describes the Cornell Massachusetts Slit
Spectrograph (CorMASS) and presents a sample of spectral
observations from its first runs on the Palomar 60-inch telescope.

CorMASS is a double-pass prism cross-dispersed NIR grating
spectrograph. Simultaneous wavelength coverage spans from
$\lambda\sim 0.75 \micron$ to the $256 \times 256$ HgCdTe
NICMOS\,3 array cut-off at $\lambda\sim 2.5 \micron$ (the
\textit{\'{z}JHK} photometric bands)\footnote{\textit{\'{z}} is a
SDSS photometric band with $\lambda_c = 0.91\micron$ and
$\Delta\lambda = 0.12\micron$.  The long wave cut-off is
determined by CCD sensitivity \citep{fuk96}.}.  This wavelength
coverage is dispersed into 6 orders on the array.  A spectral
resolution of $R\equiv\lambda/\Delta\lambda\sim 300$ maximizes
signal from relatively faint objects, while keeping resolution
high enough to identify features of the target between blended
airglow lines, including in H-band.  An externally driven cold
flip mirror permits the reflective slit to be imaged in $K_s$ on
the NICMOS\,3 array using a complementary optical
train.\footnote{$K_s$ is a passband created for the 2MASS survey
by one of us (M. Skrutskie). This band is shortened and shifted
blue-ward compared to the conventional \textit{K}-band to reduce
in-band thermal emission \citep{per98}.} The slit viewer's
$35\arcsec$ field-of-view (FOV) enables rapid object
identification and accurate placement of science targets into the
slit for spectroscopy. The compact instrument optical and
mechanical design occupies an ${\rm LN}_2$ cooled volume of
$14.5\arcsec \times 6.5\arcsec \times 6\arcsec$ inside the Dewar.

After presenting our scientific requirements for the instrument in
\S 2, the optical design is discussed in \S 3. The mechanical
design is discussed in \S 4. Electronics and Data Acquisition are
covered in \S 5.  In \S 6 we discuss performance, and in \S 7 data
reduction is discussed and sample spectra are presented.

\section{Science Requirements}
We desired an instrument that could rapidly provide identifying
spectra for candidate low-mass objects identified by 2MASS to
within 1-2 spectral sub-classes. To accomplish this the instrument
requires access to telltale L and T-dwarf identifying features in
the red-optical such as CrH and FeH at $0.86\micron$, ${\rm
H}_2{\rm O}$ at $0.94\micron$, the FeH feature at $0.99\micron$,
as well as ${\rm CH}_4$ features at $1.6\micron$ and $2.2\micron$
\citep{kir99}.  Other features important to NIR classification
include the K I resonance doublets at $1.17\micron$ and
$1.24\micron$, broad ${\rm H}_2{\rm O}$ absorption features, and
the CO bandhead at $2.29\micron$.

The instrument's resolution must be as low as possible to maximize
signal from intrinsically faint low-mass objects, yet high enough
to detect identifying features between blended airglow lines.
Since the airglow lines are most plentiful and intense in H-band,
we chose the lowest resolution that retained the ability to
observe the $1.6\micron$ methane absorption feature between
blended airglow lines, thus preserving the ability to identify
this telltale T-dwarf spectral feature for the faintest objects.

Observing efficiency was paramount if the instrument were to
quickly identify large numbers of objects and select those that
warranted higher resolution and S/N study on larger telescopes,
especially given the large number of L-dwarfs expected to be found
by surveys. This requirement favored a cross-dispersed
spectrograph with its efficient 2-D spectral format.

A NIR slit viewing mode was deemed imperative for rapid field
identification and for steering objects accurately into the slit.
Most science targets for this spectrograph have extreme visible-IR
colors, i.e. CorMASS science targets tend to be visibly faint but
IR bright.

The choice of slit width was driven by the typical seeing at the
Palomar 60-inch of $\theta_{\rm FWHM} \sim 1 \arcsec$.  To achieve
Nyquist sampling at $2 \times \theta_{\rm FWHM}$, we chose a $2
\arcsec$ slit and a plate scale of ${1 \arcsec}/\rm{pixel}$. We
desired the slit length to be as long as possible to sample the
sky for accurate subtraction ($20\arcsec$ if possible), but left
the final length to be determined by the optical design
constraints.

Lastly, we wanted the instrument to have first light within one
year of commencing design due to the rapidly progressing field of
low-mass objects.  For cost and schedule reasons we decided to
modify an existing IR Labs Dewar and use an existing NICMOS\,3
array and array electronics.

\section{Optical Design}
\subsection{Palomar 60-inch}
\begin{figure}
\figurenum{1} \label{casspix} 
\plotone{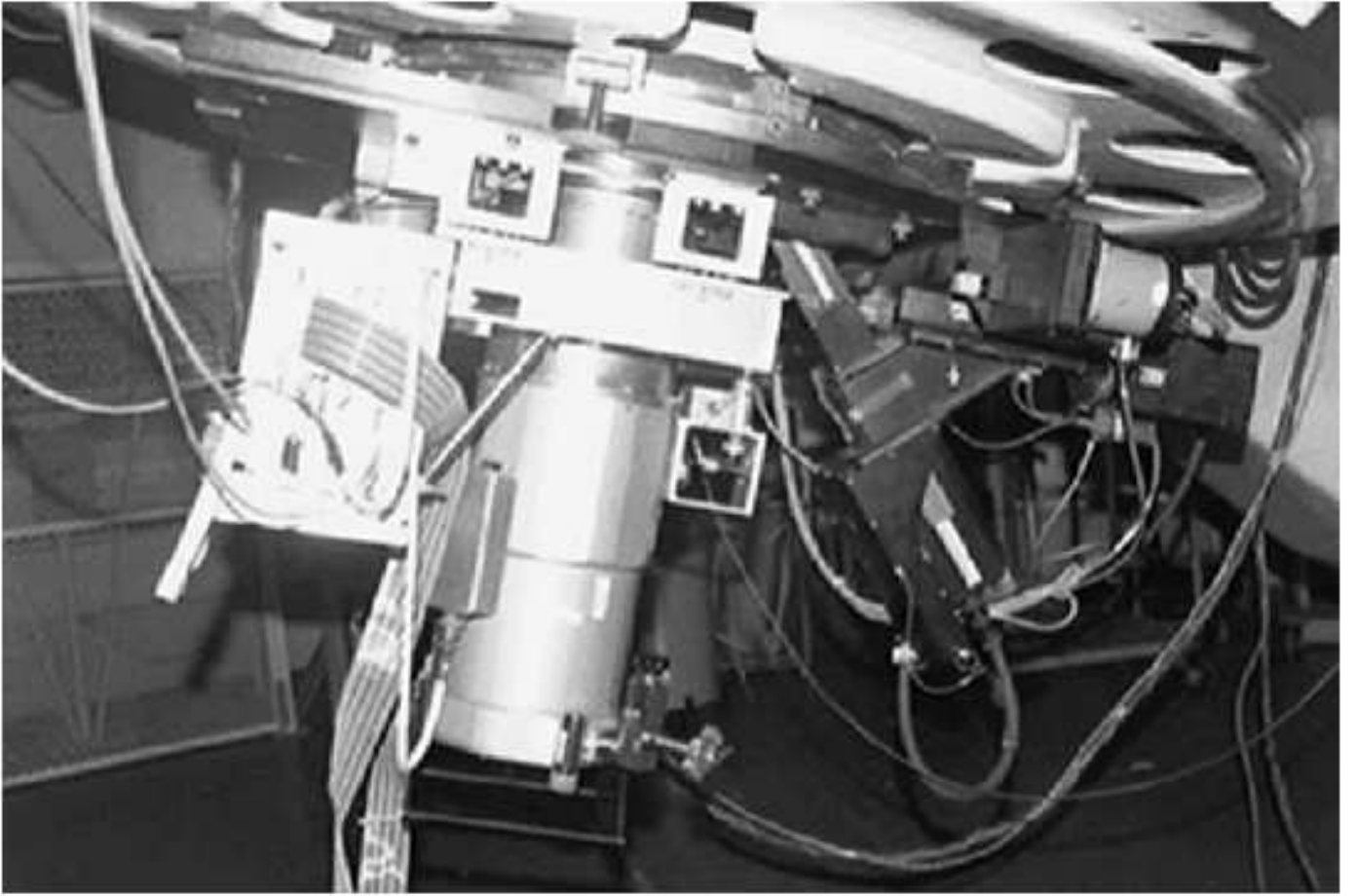}
\caption{CorMASS mounted at the Cassegrain Focus of the Palomar
60-inch telescope. On the right side of the Dewar is the
externally mounted flip-mirror drive assembly.  In the foreground
to the left is the box containing the Infrared Labs MCE-3
electronics for clocking and array readout. The telescope visual
guide camera is to the right of the Dewar mounted at $45 \degr$.}
\end{figure}

The Palomar 60-inch telescope is an equatorial mounted $f/8.75$
Ritchey-Chr\'{e}tien design \citep{bow66}.  Table \ref{tbl-sp}
includes the telescope prescription.  CorMASS mounts directly to
the instrument mounting plate at the Cassegrain focus, as shown in
Figure \ref{casspix}.  With the exception of the primary and
secondary mirrors and the Dewar window all the optical elements
are cooled to ${\rm LN}_2$ temperature.

\subsection{Design Constraints}
CorMASS features a compact design that was constrained by the
existing Dewar and the telescope focus position.

\subsubsection{Dewar Modification}
The CorMASS Dewar previously served as the 2MASS Prototype Camera
Dewar and more recently for testing NICMOS\,3 arrays for the 2MASS
survey.  The original Dewar is an Infrared Laboratories (Tucson,
AZ) ND-2(8)MOD Dewar with a 4.5 liter ${\rm LN}_2$ tank and double
shielding. It has a usable cold instrument volume of \o7.5 inches
$\times \sim5$ inches.  It was clear that a Dewar expansion was
required to increase the cold volume available for the new
spectrograph optics.  Clearly the simplest Dewar modification
would be a cylindrical extension of the Dewar walls and shields.
Thus the new spectrograph was constrained to \o $<$ 7.5 inches,
with the length a `free parameter.'

\subsubsection{Telescope Focus Position}
The optimal telescope focus is only $\sim 35$ mm below the
instrument mounting plate (Table \ref{tbl-sp}).  In addition, a
permanently mounted offset guide camera and filter motor assembly
on the telescope instrument mounting plate constrain the outer
diameter of an instrument mounted flush to the plate to be $\la
10$ inches.

These constraints on access made placement of a cold slit at the
telescope focus difficult. The awkward focus position was good
incentive to allow the beam to diverge past the telescope focus,
enter the cryogenic Dewar, and then re-image the beam onto a slit.
Such a scheme would provide straight-forward \textit{f/}conversion
between the telescope focus and slit, as well as produce a
well-defined cold pupil stop. But a re-imaging scheme would
further lengthen the cylindrical axis of the instrument, thus
increasing expense, flexure and the moment arm of the instrument
mass that must be compensated with telescope balance.

\begin{figure}
\figurenum{2} \label{benchschematic} \epsscale{0.50}
\plotone{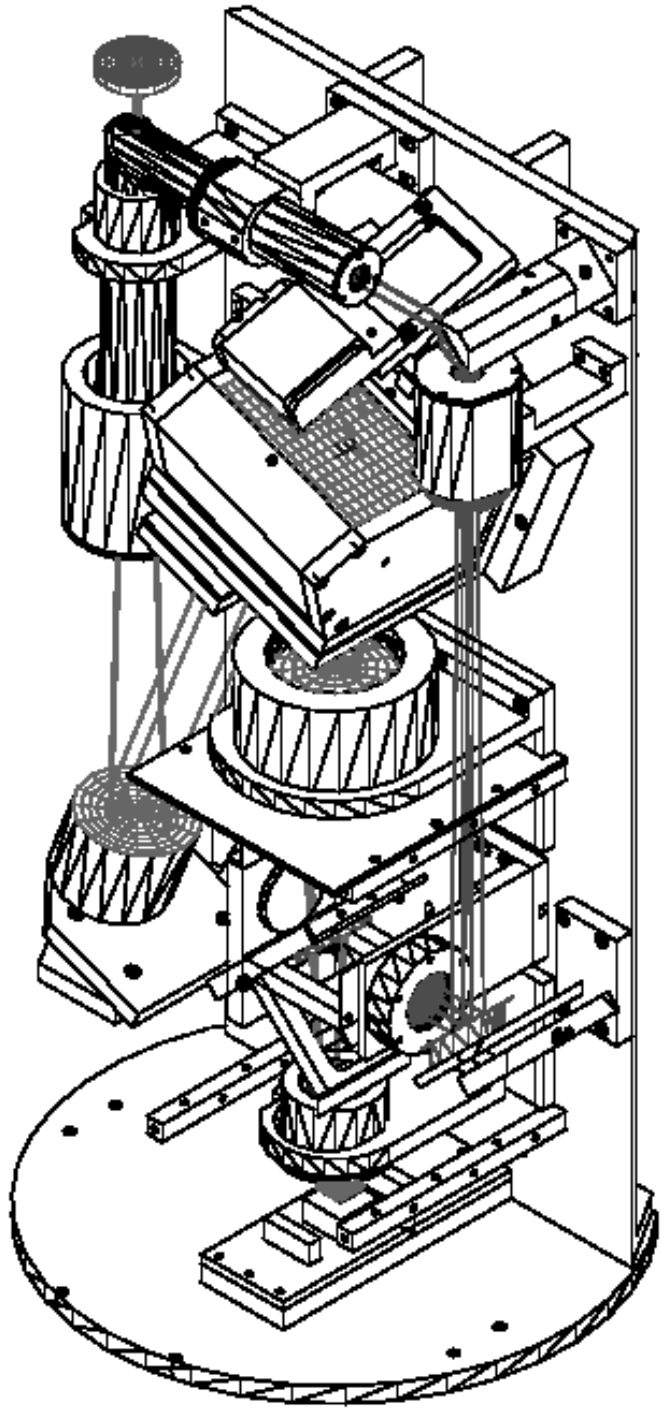} \epsscale{1.0}
\caption{The optical bench schematic. The long
dimension is $14.5\arcsec$. This bench is easily removed from the
Dewar for lab testing and alignment. The flip-mirror is shown in
both spectrograph and slit viewing positions.  The sides of the
spectrograph camera enclosure have been removed from this drawing
for display purposes.}
\end{figure}

\begin{figure}
\figurenum{3} \label{benchpix} 
\plotone{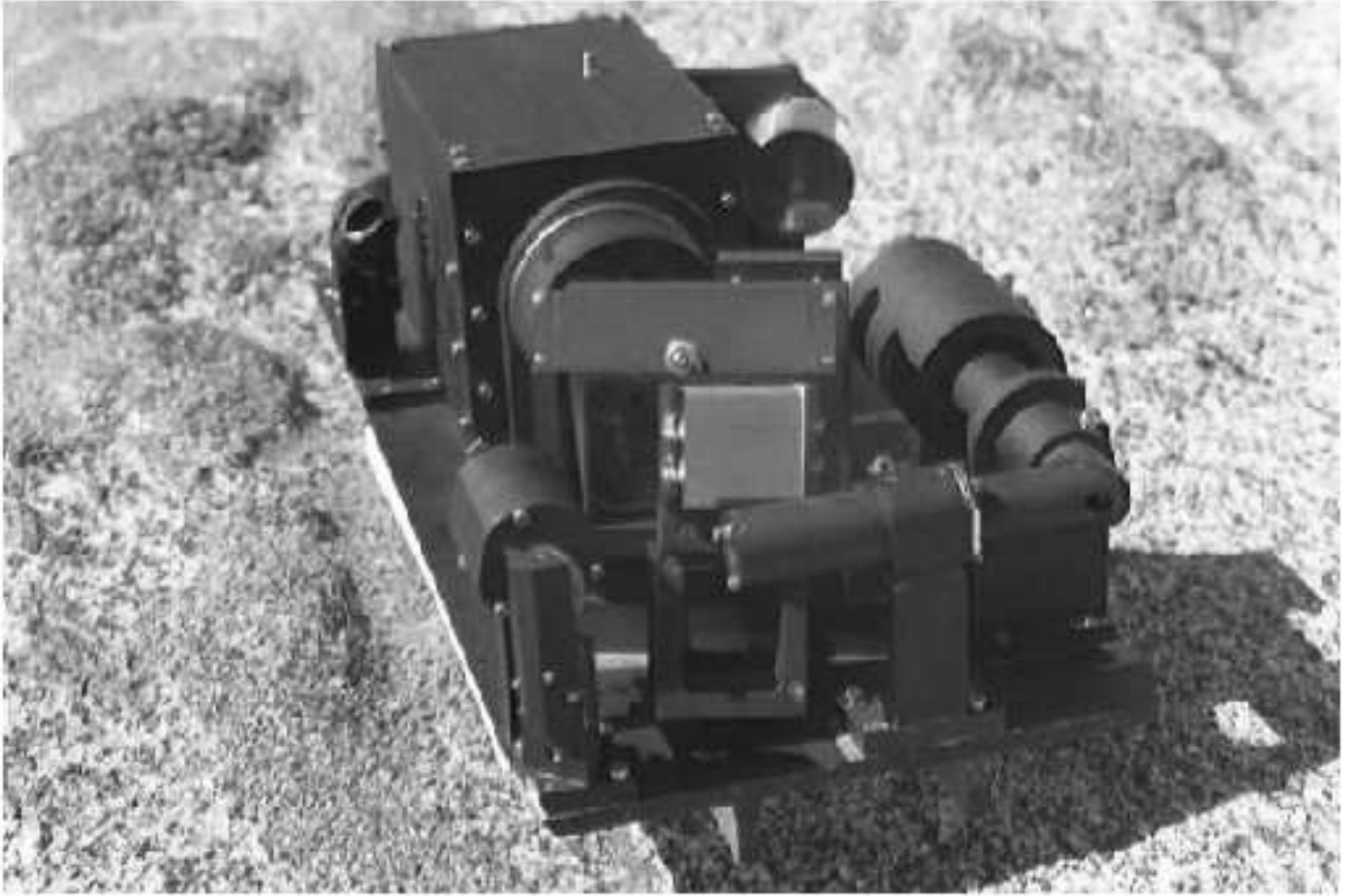} 
\caption{The
optical bench removed from the Dewar. Optical Element mount tubes
surrounding the prism are seen in the foreground and the
spectrograph camera and flip-mirror assembly is at the far end.}
\end{figure}

We chose instead the more compact scheme of placing a cold slit at
the telescope focus. This required a very tightly integrated
mechanical design just inside the window: the Dewar top plate,
CaF$_2$ window, two radiation shield tops, and slit occupy a span
of $\sim 35$ mm measured along the optical axis.
\textit{f/}conversion is accomplished as the beam diverges past
the cold slit.  The clearance between the instrument mounting
plate and the telescope fork is $38$ inches. The final design
length of CorMASS measures $27.7$ inches. Figures
\ref{benchschematic} \& \ref{benchpix} show the design of CorMASS.

\subsection{Spectrograph Optics}

\subsubsection{Prescription}
\begin{figure}
\figurenum{4} \label{opticalpath} \plotone{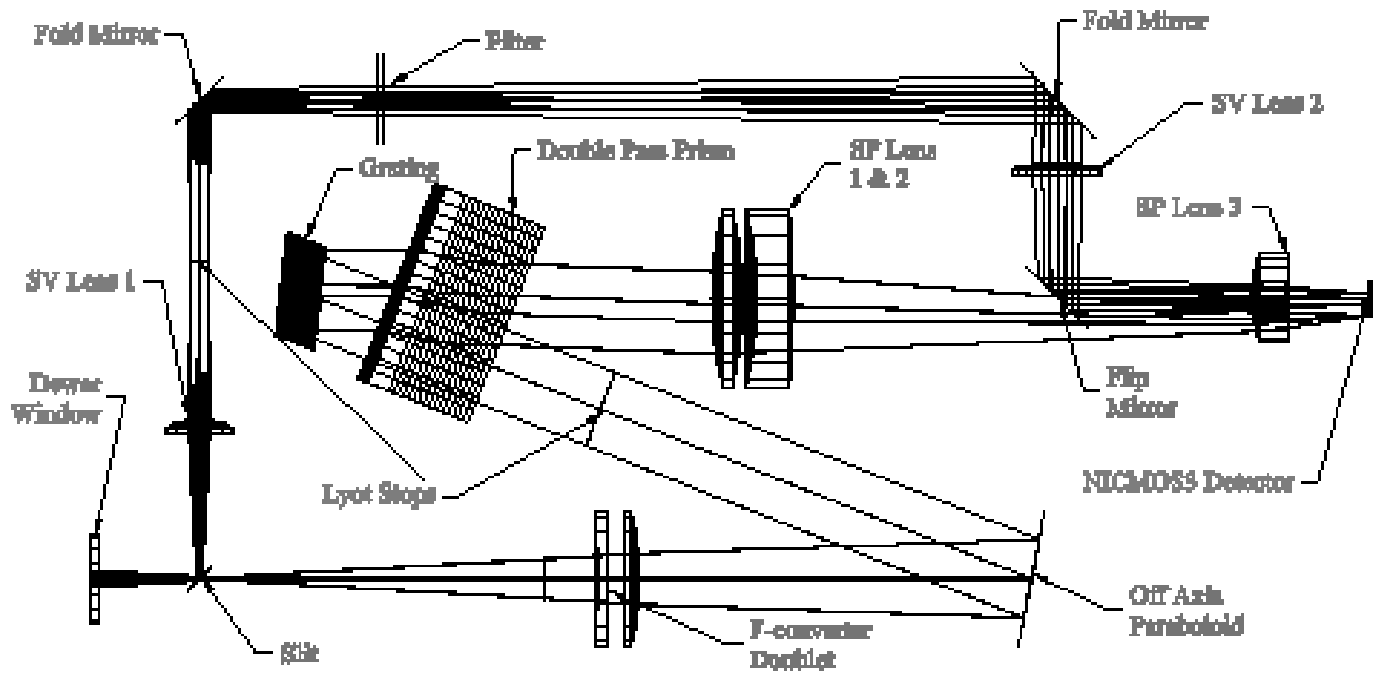}
\caption{Optical paths for the spectrograph and slit viewer. The
paths split at the slit plane. The externally driven two-position
flip mirror selects either path for imaging onto the detector.}
\end{figure}

Ray tracing, optimization, and tolerancing for the CorMASS optical
design was accomplished with the optical design software ZEMAX
Version 6.0 (Focusoft, Tucson, AZ). The optical layout of CorMASS
is shown in Figure \ref{opticalpath}.  All five lenses are
spherical and one is from stock while the remaining four are
custom manufactured using in-house test-plates by Janos
Technology, Inc. (Townshend, VT).

The main dispersing element is a coarse 40 lines mm$^{-1}$
grating, blazed at $4.8\micron$.  A prism made from the Schott
Dense Flint SF-10 is used in double-pass for cross-dispersion.
Table \ref{tbl-sp} lists the spectrograph prescription.

After passing through the ${\rm CaF}_2$ Dewar window, the
converging telescope beam reaches a focus at the slit. A wire
electrical discharge machine (EDM) was used to cut a $185 \micron
\times 951 \micron$ slit into the center of a $.008\arcsec$ thick
stainless steel (SS304) disk that had been hand polished with $1
\micron$ diamond paper.  Ron Witherspoon Inc. (Campbell, CA)
manufactured the slit.  The disk is rotated $45 \deg$ about the
axis of the slit length so that incoming light not transmitted
through the slit is reflected into the slit viewing train. The
slit width is oversized by $\sqrt{2}$ to compensate for the disk
rotation.

Past the slit the beam is converted from $f/8.75$ to $\sim f/14$
by an Infrasil 301-CaF$_2$ air-spaced doublet\footnote{Infrasil
301 is a water-free fused quartz made by Hereaus-Amersil Corp.
(Duluth, GA).} and is collimated by a gold-coated, diamond-turned,
aluminum alloy (6061-T6) off-axis paraboloid. The beam then passes
through a cold Lyot Stop sized to the pupil diameter of 22.6 mm.
After the stop the beam passes through the SF-10 prism for the
first cross-dispersion pass.  The prism has a $30 \deg$ apex angle
and is used at minimum deviation.  It was fabricated by Kreischer
Optics, Ltd. (McHenry, IL).

A prism was necessary as the cross-dispersing element because the
desired wavelength coverage exceeded a factor of 2.  A double-pass
prism design was adopted because it could be implemented in
combination with the grating to produce a very compact optical
package.  When the dispersing elements were ordered in series \it
[prism pass 1 (cross dispersion) -- grating reflection (primary
dispersion) -- prism pass 2 (cross dispersion)]\rm \, the optical
path was neatly returned into the main `2-d' plane of the optical
bench (Figures \ref{benchschematic} - \ref{opticalpath}).
Single-pass prism designs led to cubical mechanical packages that
could not be simply inserted into the dewar.

In addition to SF-10 various other prism materials were
considered, including AMTIR1 and ZnSe.  Both have higher
dispersion in the NIR than SF-10 but were rejected due to coating
difficulty and expense. The use of a material easily coated was
imperative for the reduction of Fresnel reflection losses at the
four vacuum-prism interfaces encountered with a prism used in
double-pass.\footnote{We became aware of SF-10 during a search of
stock prism vendors. Interestingly, the Schott `SF' dense flint
glasses have a partial dispersion similar to the alkaline-earth
flouride crystals (e.g. ${\rm BaF}_2$, ${\rm SrF}_2$, and ${\rm
CaF}_2$). This makes the SF glasses an appealing `flint' to couple
with the crystals to form achromatic NIR cameras \citep{oli97}.}
The dispersion of SF-10 and the resulting order separation is
discussed in \S \ref{order-format}.

\begin{figure}
\figurenum{5}
\label{SF10trans}
\plotone{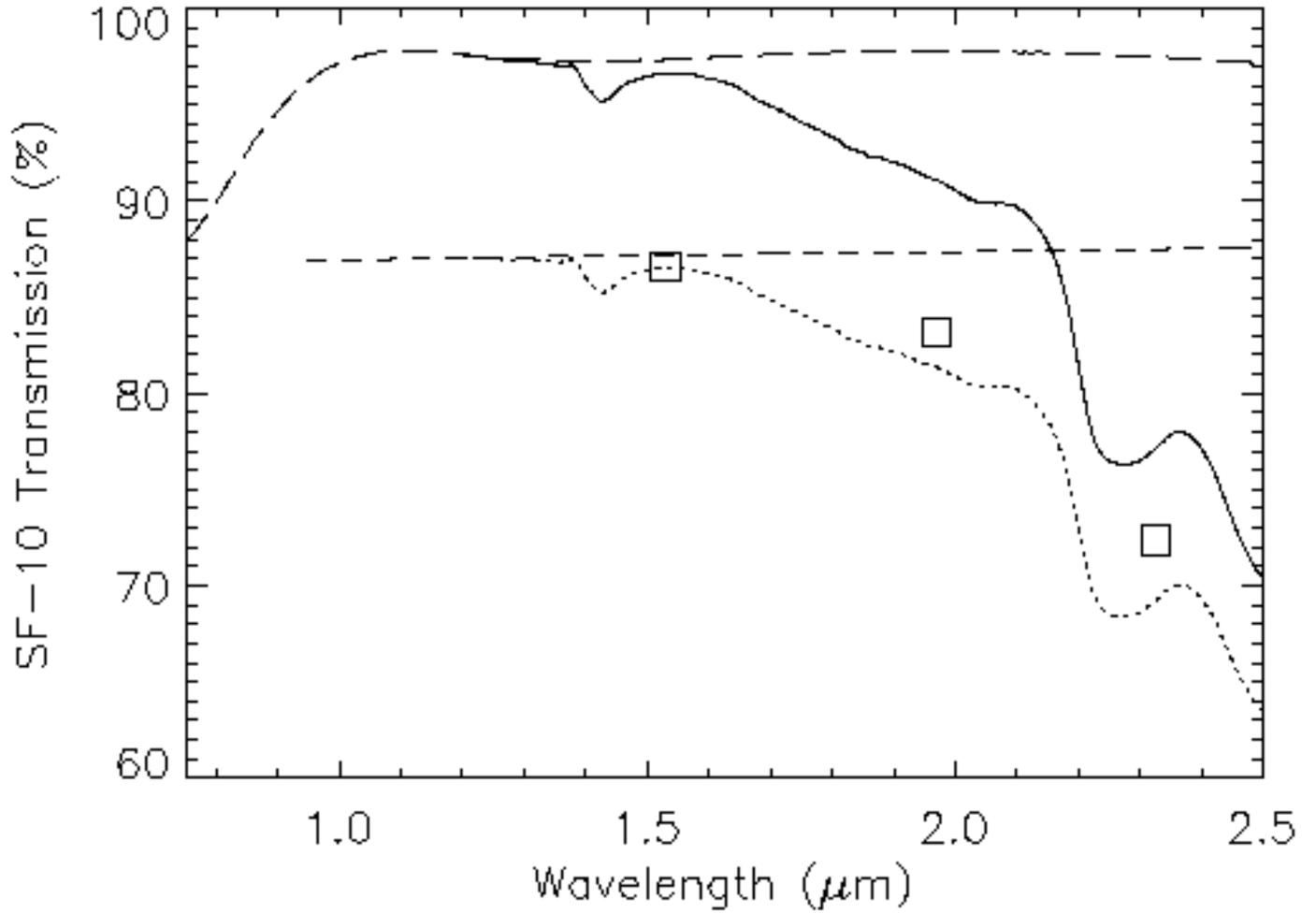}
\caption{Measured internal transmission of SF-10 (21.7
mm thickness) (dotted line) including theoretical reflection
losses from two air-glass interfaces (dashed line).  Square boxes
represent absorption prediction derived from Schott Data Sheet for
uncoated material.  Expected AR coating performance for two
air-glass interfaces (long dashed line) is applied to the measured
internal transmission to estimate total transmission for one prism
pass in CorMASS (solid line).}
\end{figure}

The measured internal transmission of SF-10 (kindly provided by
Dr. E. Oliva) and expected Anti-Reflection (AR) coating
performance (Janos Technology, Inc.) are plotted in Figure
\ref{SF10trans}. The transmission measurement was performed as
described in \citet{oli97} and includes theoretical Fresnel losses
from two air-glass interfaces (dashed line) and internal
transmission through one pass of a 21.7 mm uncoated sample (dotted
line).  This is very close to the median length of SF-10 traversed
by the CorMASS beam in one pass (20.7 mm), so the plotted
transmission, neglecting Fresnel losses, accurately reflects the
expected internal transmission in our instrument. Shortward of
$1.2\micron$ the glass is nearly transparent.  We assume
absorption features longward of $1.4\micron$ are due to water/OH
in the glass. Combining the expected reflection losses through two
air-glass interfaces after AR coating (long dashed line) with the
internal transmission gives the total expected transmission
through one prism pass for CorMASS (solid line). While the
transmission losses were unfortunate, they were accepted because
the primary spectral features in $K$-band for L and T-dwarfs, the
CO bandhead at $2.29\micron$ and the $2.2\micron$ ${\rm CH}_4$
feature, resp., were deemed less important and redundant compared
to the \textit{\'{z}JH} band features.

Following the first pass through the prism the beam is diffracted
by the grating. Richardson Grating Labs (Rochester, NY) mounted an
epoxy replica of a stock master grating onto an aluminum alloy
(6061-T6) substrate provided by us. The gold coated grating, with
40 ${\rm lines}/{\rm mm}$ groove density and blazed at $5.5\degr$,
is used with a turn-angle of $21.5\degr$ to match the angular beam
deviation of the off-axis paraboloid for optical layout symmetry.
The first order blaze wavelength ($\lambda_{\rm{B}}$) is
$4.71\micron$. The detector images the 2nd through 9th orders,
with coverage from $2.5\micron$ to $\rm{H}\alpha (0.6563\micron)$.
We designed the spectrograph to use orders 2-5 and were fortunate
that orders 6-7 are in good optical focus as well. In practice
only 2nd through 6th orders have useful throughput. (Spectrograph
throughput is discussed in \S 6).

Inadvertently the shortward edge of order 2 and longward edge of
order 3 are cutoff by the array edges, resulting in a gap in
spectral coverage from $2.014$ - $2.035 \micron$.  Evidently the
final spectrograph camera focal length is too long, likely due to
errors in our assumed cryogenic indices of refraction.

Grating efficiency is well described by scalar theory if $\lambda
/d < 0.2$ where $d$ is the groove spacing ($25\micron$ for
CorMASS).  Therefore, polarization effects are very small
\citep{loe77}. Theory also predicts $\ga 50\%$ efficiency from
$2/3 \lambda_{\rm{B}}$ to $1.8 \lambda_{\rm{B}}$ in first order.
At higher orders this efficiency will be maintained because the
grating is used even farther into the scalar domain as $\lambda/d$
decreases \citep{loe77}.  A check of our chosen grating parameters
in a commercial software program for grating efficiency analysis
(PC Grate, Optometrics, Ayer, MA) yields estimated peak
efficiencies of 90\% for orders 2-5.

After diffraction by the grating, the beam passes through the
prism for a second time.  Lastly, the beam is focused onto the
detector by an $f/5.5$ camera which delivers a final plate scale
of $1\arcsec/\rm{pixel}$ to the $40 \micron$ square NICMOS\,3
pixels. The camera features three elements in a doublet-singlet
format. Following a BaF$_2$-Infrasil air-spaced doublet, there is
a space of $\sim 5$ inches before the converging beam passes
through the final Infrasil lens and focuses on the detector.  The
final lens and detector are separated by $\sim 1$ inch.

\begin{figure}
\figurenum{6}
\label{ee}
\plotone{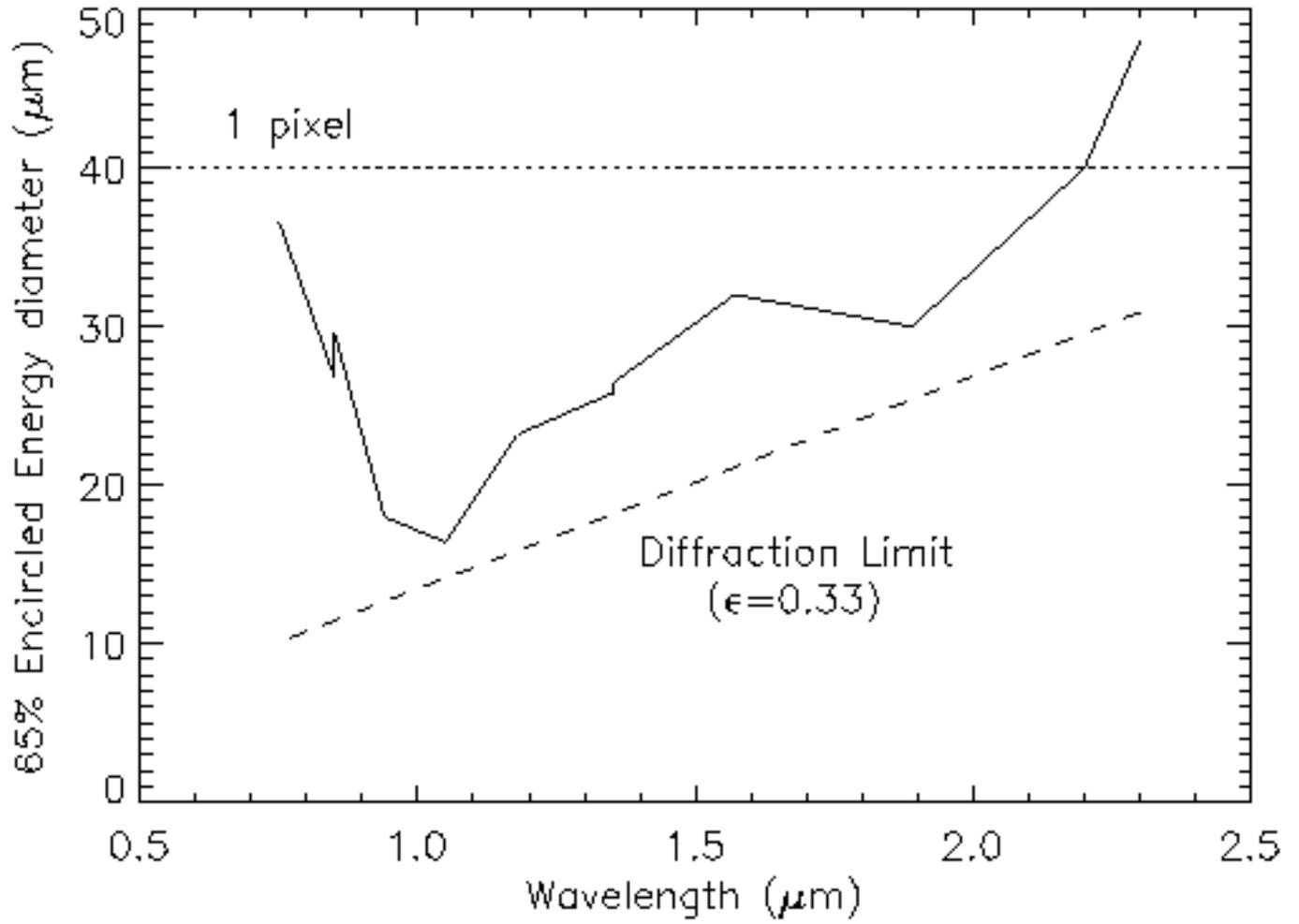}
\caption{The 65\% Encircled Energy diameter  v.
wavelength for the spectrograph (entire slit FOV). The approximate
diffraction limit, dashed line, and the pixel size, dotted line,
are also shown.}
\end{figure}

The ratio of the telescope central obscuration diameter to primary
diameter is $\epsilon = 0.3$.  For a telescope with $\epsilon =
0.33$ the Encircled Energy fraction (EE) within the first Airy
dark ring is 0.65 \citep{sch87}.  We used this EE benchmark to
judge the image quality of the spectrograph design.  Shortward of
$2.2 \micron$ all wavelengths image EE $\geq 0.65$ within 1 pixel
(Figure \ref{ee}).  This applies for the entire slit FOV (2\arcsec
x 15\arcsec).

\subsubsection{Order Format}\label{order-format}
\begin{figure}
\figurenum{7} \label{fpaimages} \plotone{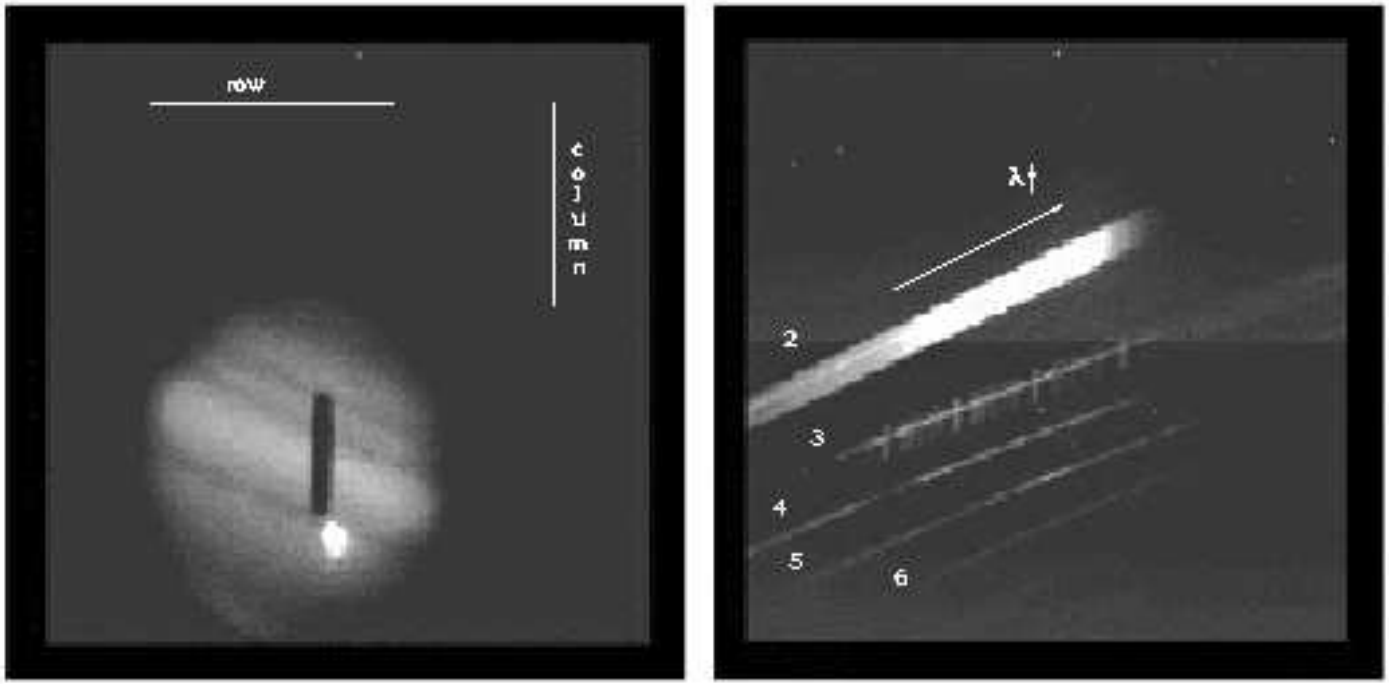} \caption{Slit
Viewer and Spectrograph images. (a) Jupiter fills the $35\arcsec$
Slit Viewer FOV, with Io just below the $2\arcsec \times
15\arcsec$ slit. Row and column directions are indicated.  (b) The
spectral image displays a Wolf-Rayet near the middle of the slit.
The second order is at the top, sixth order at the bottom. The
thermal emission of the atmosphere increases rapidly in the second
order then abruptly cuts off at $\sim 2.5\micron$. Sky lines
dominate the third order (H-band). NICMOS\,3 shading anomaly is
seen at the bottom of the quadrants prior to correction.  Order
numbers and direction of increasing wavelength are indicated.}
\end{figure}

\begin{figure}
\figurenum{8}
\label{dispersionfig}
\plottwo{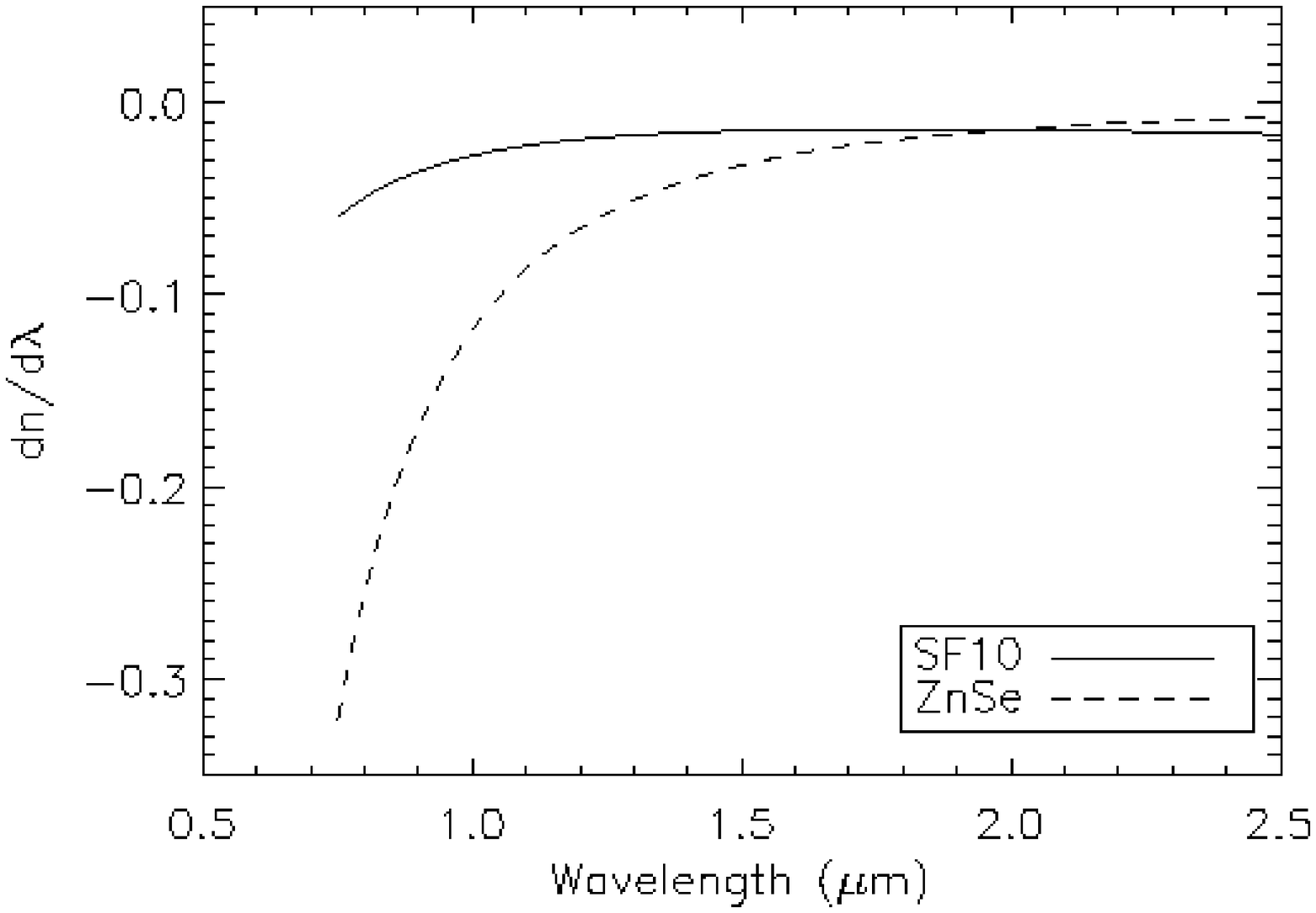}{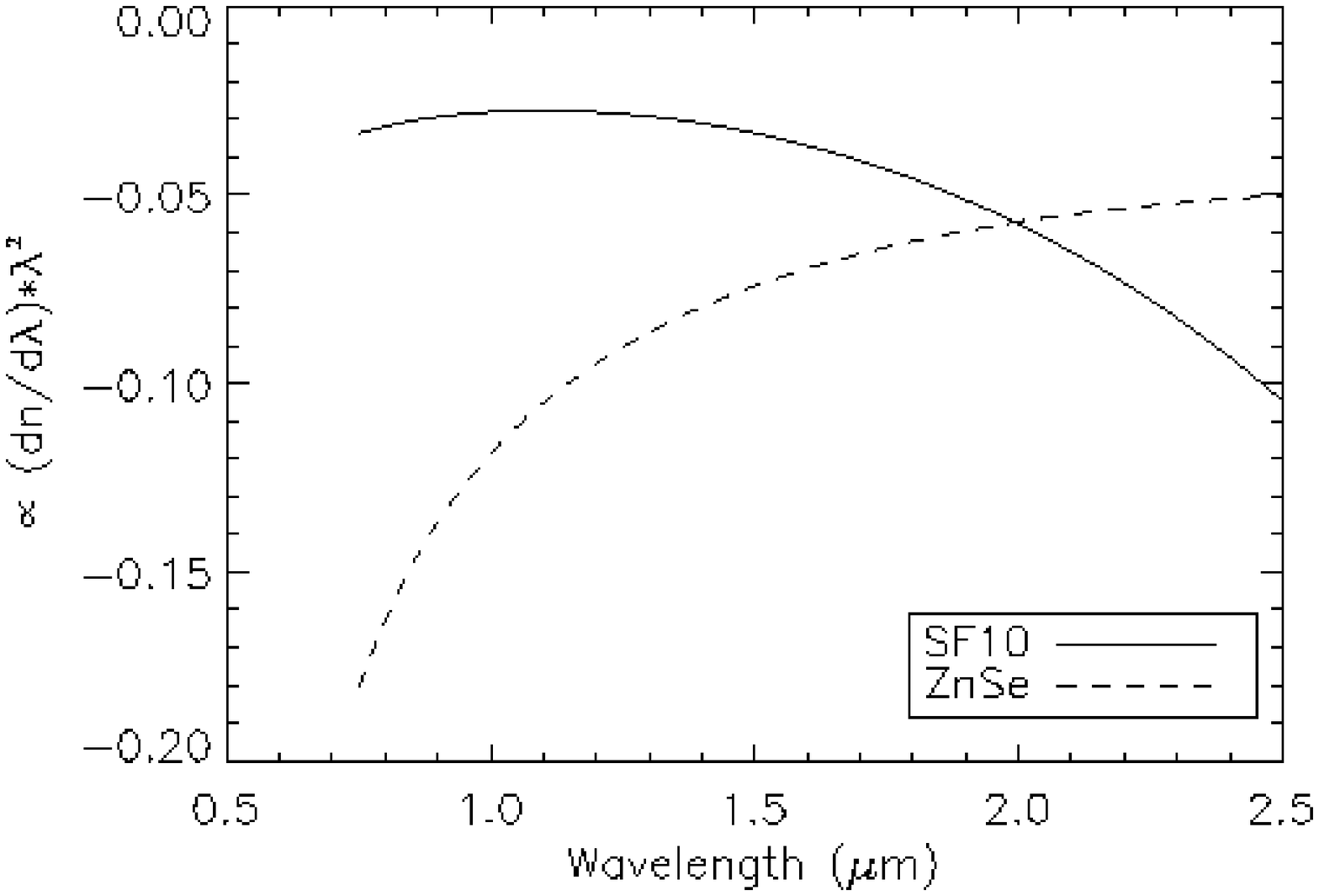}
\caption{(a) Dispersion ($dn/d\lambda$) vs. $\lambda$
and (b) Relative order separation vs. $\lambda$ for Schott dense
flint SF-10 and ZnSe at room temperature. SF-10 is used for the
cross-dispersing double pass prism.}
\end{figure}

The SF-10 prism in CorMASS leads to an order format similar to
that found when a first order grating is used as the
cross-disperser\footnote{Grating cross-dispersers usually give an
order separation $\propto \lambda^{2}$ \citep{sch87}.}; CorMASS'
order separation increases with increasing wavelength (Figure
\ref{fpaimages}), whereas prism cross-dispersers typically give
larger order separation for decreasing wavelength.  The CorMASS
format occurs because the dispersion of SF-10 in the near-infrared
does not follow the usual dispersion law for transparent glasses
in the optical, $dn / d\lambda \propto \lambda^{-3}$. Instead of a
monotonically decreasing dispersion with increasing wavelength,
SF-10 has a dispersion minimum at $\sim 1.7\micron$ (Figure
\ref{dispersionfig}) at room temperature. This leads to
theoretical relative order separations also shown in Figure
\ref{dispersionfig}. Both plots compare SF-10 and ZnSe.  SF-10 has
a similar dispersion function at 77K as at room temperature.

\subsection{Slit Viewer Optics}
Complementing the spectrograph optical train is a slit viewing
mode (Figure \ref{opticalpath}).  Light not transmitted through
the slit is reflected by the polished slit substrate into the
secondary slit viewing train. This allows viewing a $K_s$ image of
the reflective slit for rapidly steering science targets into the
slit.  A remotely-controlled drive motor outside the Dewar turns
an internal flip-mirror between the slit-viewing and
spectrographic positions in $\sim 10$ sec. The flip mirror is
located between the doublet and singlet of the spectrograph
camera.  The slit viewing optical prescription is listed in Table
\ref{tbl-sv}. Since the slit viewing mode is not intended for
photometric imaging, all lenses in the slit viewing train are from
manufacturer's stock to reduce material costs.

Light reflected from the slit is collimated by a CaF$_2$ lens and
passes through a Lyot Stop sized to the pupil diameter of 5.1 mm.
The collimated beam is then steered by two gold-coated fold
mirrors to the spectrograph camera assembly.  The beam also passes
through a $K_s$ filter mounted between the fold mirrors (tilted
$5\degr$ away from the spectrograph elements to reduce light
contamination). The slit viewing filter can be changed by opening
the dewar. Lastly the beam is refocused onto the detector by the
combined action of a CaF$_2$ lens and the final Infrasil lens of
the spectrograph camera (Figure \ref{opticalpath}).  The
gold-coated flip mirror steers the light from the CaF$_2$ lens
towards the final Infrasil lens and detector.  The plate scale of
the slit viewer is $0.27\arcsec/\rm{pixel}$, corresponding to an
$f/18.85$ beam. This plate scale was chosen to give a FOV
($35\arcsec$) large enough to aid the identification of targets in
confused fields (Figure \ref{fpaimages}).  The slit-viewer design
gives 65\% encircled energy diameters of $\leq 3$ pixels on axis
and $\leq 4$ pixels at the edge of the FOV.

Light leakage from the slit viewing train when in spectroscopy
mode is minimal because the flip mirror mechanism blocks the slit
viewing beam just past the second slit viewing lens with a shaped
baffle plate when not in use (Figure \ref{benchschematic}). We
gave extra care to minimize this source of stray light in such
close proximity to the detector.

\section{Mechanical Design \& Alignment}
\subsection{Mechanical Design}
AutoCAD Version 14.0 and Mechanical Desktop 2.0 (Autodesk, San
Rafael, CA) were used for the mechanical design.  The optics were
designed assuming all elements inside the Dewar are at 77K.  Since
the optical mounts are machined at room temperature, distances
between elements required by the optical prescription are adjusted
for the linear expansion of materials (Table \ref{tbl-cle}) prior
to the mechanical design and fabrication. Thus upon cooling the
elements are correctly positioned.

All optical mounts were fabricated from the aluminum alloy
6061-T6, the same material used for the off-axis paraboloid and
grating substrate.  The mounts are attached to a `T' cross-section
aluminum `optical bench' extending from the Dewar's cold plate.
The use of only one material for all mounts reduced Coefficient of
Thermal Expansion (CTE) effects and eased the difficulty of
achieving good optical alignment at 77K.  The aluminum optical
bench, off-axis paraboloid, and grating substrate were thermally
cycled between room temperature and 77K multiple times prior to
final finishing to stabilize them.

\begin{figure}
\figurenum{9} \label{prismcage}
\plottwo{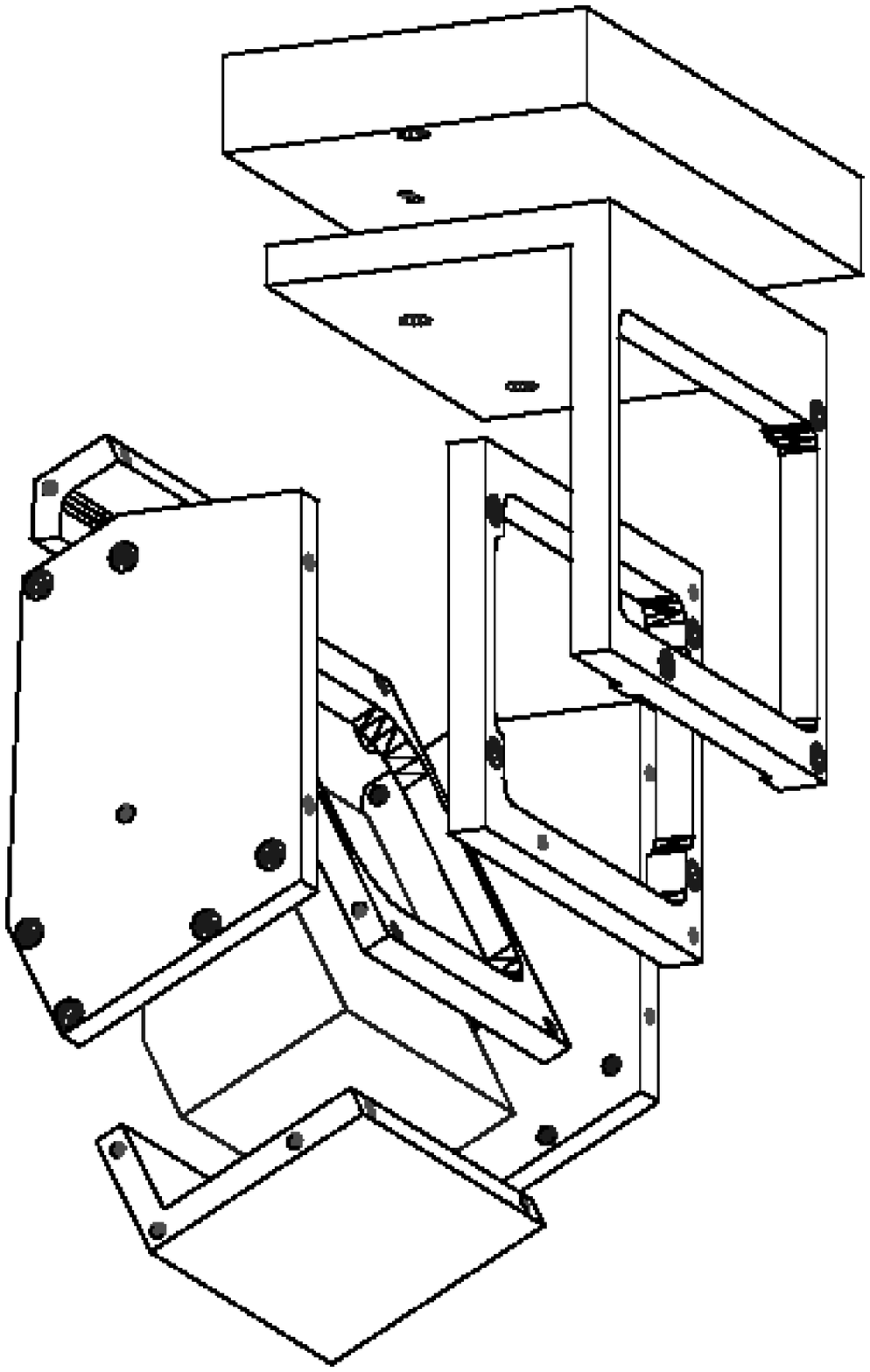}{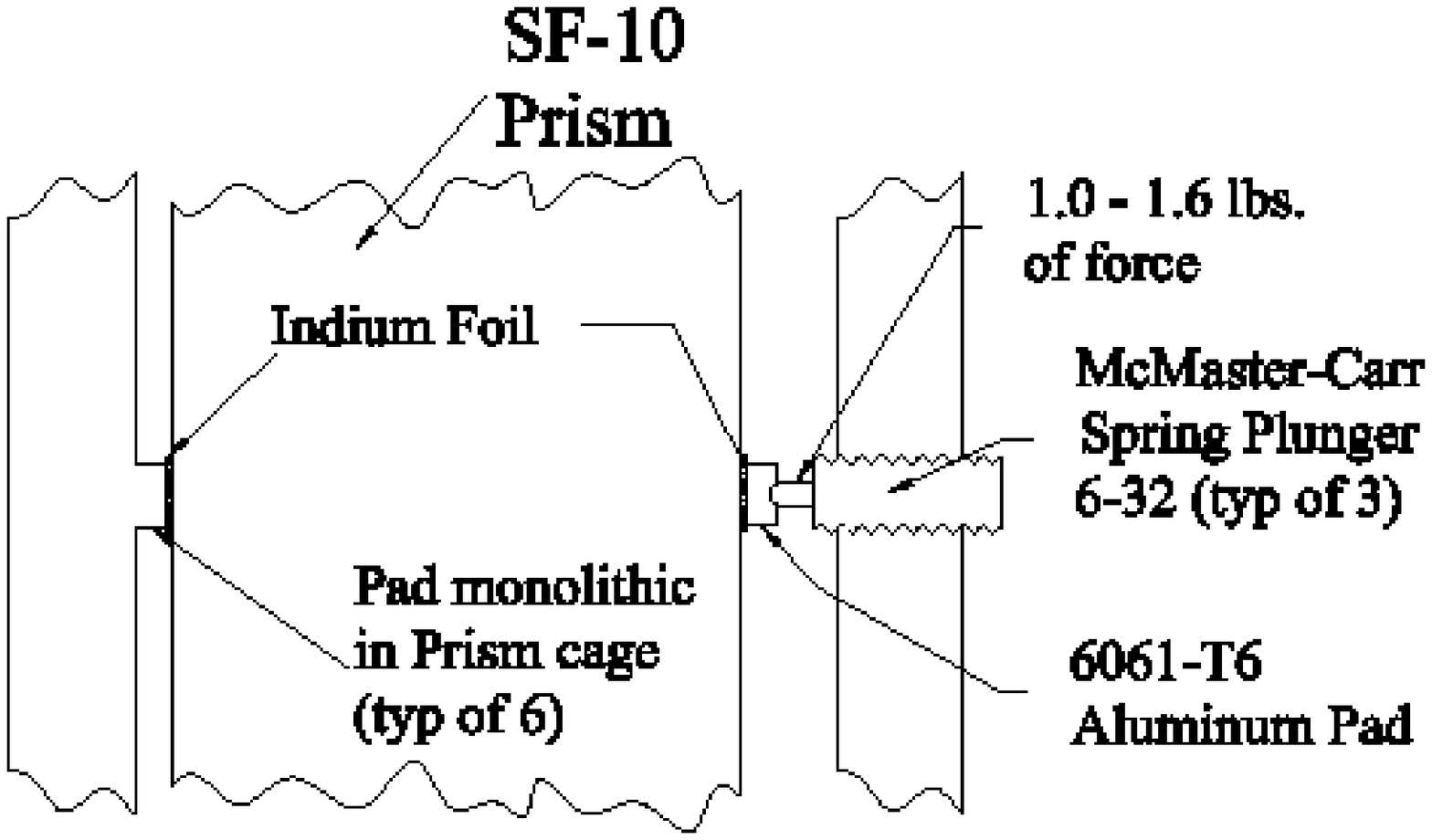}
\caption{(a) The prism cage constrains the prism in a
semi-kinematic mount and provides stray-light control. (b) The
prism holding scheme includes three spring plungers that oppose
six monolithic pads in 3-dimensions to secure the prism and
compensate for thermal expansion effects.}
\end{figure}

The prism is constrained through a system of six aluminum pads
that form a semi-kinematic mount (Figure \ref{prismcage}).  Three
spring plungers oppose the pads in three dimensions and each
exerts $1.0-1.6$ lbs force to constrain the $\sim 0.7$ pound
prism. Between each pad and the prism material is an Indium shim
to prevent abrasion (Figure \ref{prismcage}).

To accommodate differing thermal contractions all lenses are
mounted in lens barrels with oversized inside diameters to prevent
crushing.  Retaining rings compress spring wave washers against
the outer annulus of lens faces to accommodate differences in
axial contraction during cooldowns.

The flip mirror is driven by an externally mounted 30 VDC motor. A
shaft directly couples the external motor to the flip mirror
through an O-ring rotary seal at the Dewar wall.  The O-ring seal
and a flexible coupling inside the dewar make up for small
positional and axial changes upon cooling the instrument.  Mirror
position is controlled by a logic circuit involving a remotely
operated toggle switch at the observer's station and limit
switches engaged by the motor shaft outside of the Dewar.
Placement of the limit switches outside the Dewar allows
adjustment of the mirror-throw while the instrument is cold.

The double-shielded dewar uses only $\sim 1$ liter of ${\rm LN}_2$
per night of observing.

\subsection{Stray Light Reduction}
All appropriate surfaces are painted black for stray light
reduction. After using one primer coat of the two-part 9924 Wash
Primer, black polyurethane Aeroglaze Z-306 paint (Lord
Corporation, Erie, PA) mixed with carbon black powder was applied
in two or more coatings to all exposed aluminum surfaces
\citep{hun96}.

The optical element mounts were designed to minimize stray light.
Lenses are mounted within cylindrical aluminum tubes, as mentioned
above. Particular attention was devoted to reducing stray light
adjacent to the prism.  Since the reflective slit is close to the
prism, both optical paths past the slit are well enclosed in
tubes.  The prism itself is enclosed in a mount that exposes only
the entrance and exit faces to the optical beam. Another critical
area was that near the detector.  To reduce stray light the final
spectrograph camera and flip mirror assembly were placed within a
rectangular enclosure that extends nearly to the cold plate to
prevent the direct illumination of the detector by stray light
from the shield walls (Figure \ref{benchpix}). All transmissive
elements in the spectrograph train were AR coated.

\subsection{Alignment}
Alignment of the optical trains was completed in two parts -
internal element alignment and alignment with the telescope
secondary. First, alignment within the instrument was conducted in
the lab with a HeNe laser. As discussed above, because all element
mounts were made from aluminum, the relative alignment of the
optical elements was preserved upon cooling from room temperature
to 77K. A telescope simulator was used to re-image a neon
illuminated point source onto the slit plane to check focus. The
slit-viewing collimating lens and spectrograph f/conversion lens
positions were adjusted to make the slit-viewer field and the
spectrum confocal with the slit.

Second, the independent optical paths of the spectrograph and slit
viewer were co-aligned with the telescope secondary through an
iterative process.  Using a gimbaled sighting scope mounted on the
telescope at the same position as CorMASS normally mounts, the
scope was centered on the center of the secondary. The scope (with
its directional alignment preserved) was transferred to a
cassegrain-mounting-plate-simulator in the lab and sighted onto a
paper disk the diameter of the secondary and mounted at the
distance of the secondary. CorMASS was then mounted onto the
simulator and a neon lamp was used to trace the \textit{f/}cone of
both optical paths at the distance of a paper disk.  Initially,
the spectrograph \textit{f/}cone was found to be aligned within
$0.5\degr$ of the secondary (centers within 1 inch at the
slit-secondary distance of 135 inches), but the slit viewing
\textit{f/}cone was found to be displaced 4 inches. The slit
viewing train elements were re-aligned and a subsequent test
showed the alignment to have improved to within $0.5\degr$.

Peak-up of both trains relative to the secondary can be fine-tuned
by adjusting the position of 4 mounting blocks that attach the
instrument to the Cassegrain mounting plate.

\section{Electronics \& Data Acquisition}
A set of MCE-3 electronics from Infrared Laboratories provides
clocking and readout of the NICMOS\,3 array.  Four 16-bit
analog-to-digital converters sample the output of each of the four
NICMOS\,3 quadrants with a pixel dwell time of 3 $\mu {\rm sec}$.
A complete read of the array requires 51 $m {\rm sec}$. Data are
recorded in a standard double-correlated sampling sequence: reset,
${\rm read_1}$, integration delay ($t_{int}$), ${\rm read_2}$. The
reset timing is identical to the read timing with a pixel dwell
time of 3 $\mu {\rm sec}$.  A fiber-optic interface transmits the
sampled data from the telescope to the control room where a custom
interface provides for buffering of the frames into a
frame-grabber card in a Pentium-based data-acquisition computer.
The computer stores the raw reads as two-plane FITS images and
displays the ${\rm read_2}-{\rm read_1}$ frame difference for
real-time analysis by the observer.

A custom DOS-based user interface provides a variety of control
features such as background subtraction of stored images and
Ethernet connection to the Telescope Control System (TCS). There
are two acquisition modes: continuous-acquisition and single frame
modes. The former provides a real-time stream of typically short
integration frames ($\geq 0.5$ sec) useful for steering objects
into the slit.

\section{Performance}
\begin{figure}
\figurenum{10}
\label{throughput}
\plotone{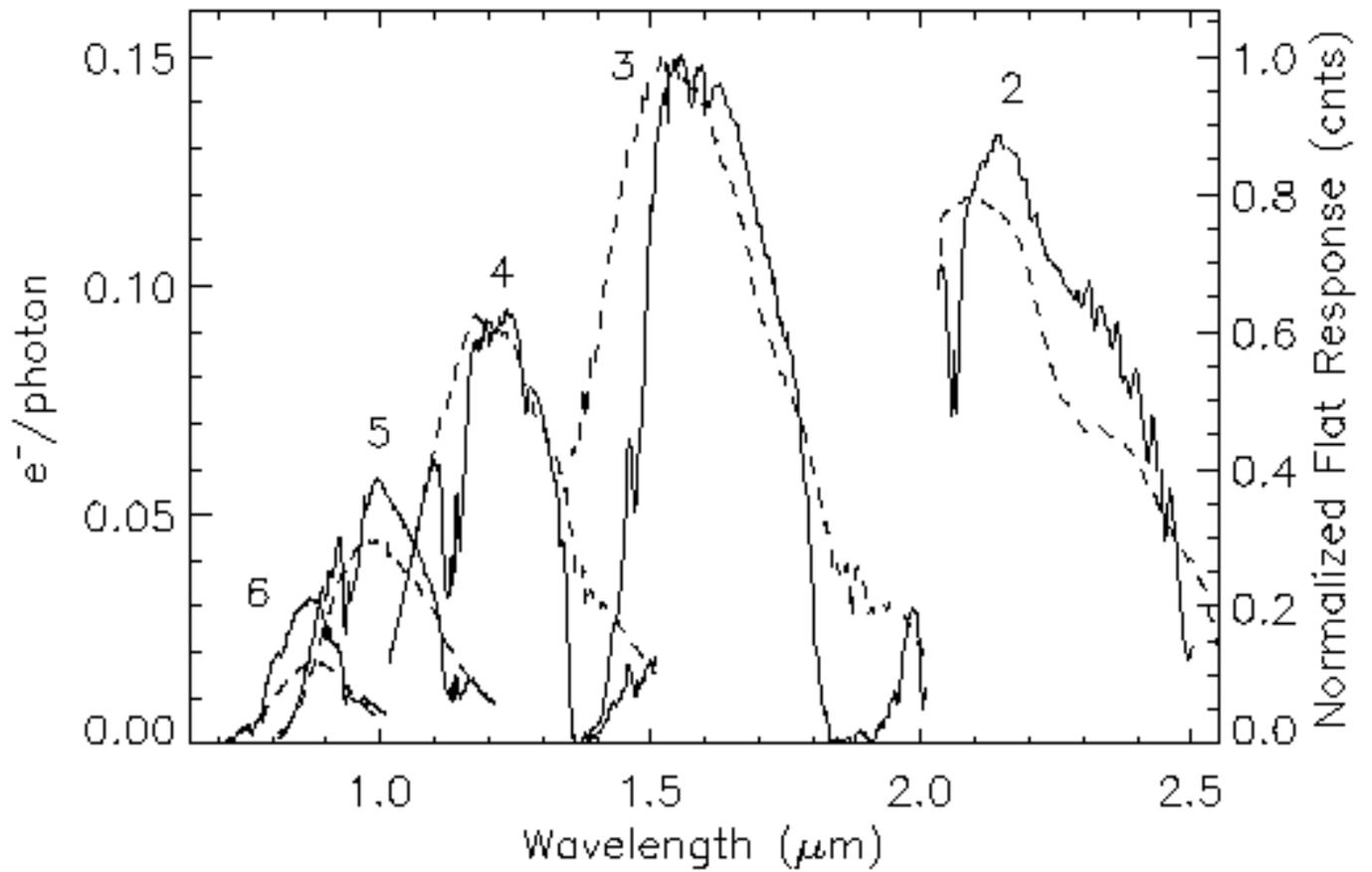}
\caption{Spectrograph throughput v. wavelength for
each order (solid line). Order number is identified at the top of
each throughput curve. Throughput is in electrons out of the
detector per photons in at the top of the atmosphere.  Also
plotted for comparison is the dome flat response in raw counts
normalized to the peak in $H$ (dashed line).}
\end{figure}

Detector and instrument performance are summarized in Table
\ref{performance}.  Average spectrograph throughput is
$\textit{\'{z}}$=0.05, $J=0.07$, $H=0.13$, and $K = 0.10$. Figure
\ref{throughput} gives throughput as a function of wavelength and
order, and includes transmission losses from the atmosphere,
telescope, instrument and detector.  The spectrograph sensitivity
is $J=14.6$, $H=14.9$, and $K=14.0$ in $t_{int}=3600$ sec for
$S/N=5$.  The spectrograph performance was determined by analyzing
the reduced spectrum of the Faint HST standard P565-C
\citep{per98} observed in clear conditions and with seeing $\la 2
\arcsec$ at $K$.  This sensitivity conforms to experience; in
clear conditions we normally observe $K_s \sim 13$ objects for
$t_{int} \sim 3000$ sec, and $K_s \sim 14$ objects for $t_{int}
\sim 4000$ sec, to ensure sufficient S/N.

Also plotted in Figure \ref{throughput} are normalized dome flat
response curves for each order.  The response curves displayed are
the normalized spectra (raw DN) from the sum of 45 \it flat lamp
on -- lamp off \rm \, differenced pairs accumulated over 9 nights
in May 2000.

The Slit Viewer sensitivity is $K_s \sim 13$ in 2 sec for a
$\sigma = 5$ point source detection with a subtracted background
under good seeing conditions.  In practice $t_{int}$ $\sim$ 2-5
sec continuous acquisition frames without background subtraction
are used to detect and steer $K_s \leq 13$ sources into the slit.
For $K_s = 14$ (the faintest sources practical to observe
spectroscopically with CorMASS) $t_{int}$ $\sim 20$ sec frames are
used. Subtraction of a previously saved sky background integration
is often used to aid detection of fainter sources.  Slit Viewer
throughput at $K_s$ is $\sim 0.20$ $e^{-}$ out of the array per
photon in at the top of the atmosphere.  This throughput includes
transmission losses from the atmosphere, telescope, instrument
optics, and detector QE.  The slit viewer performance was
determined by analyzing images of P-565C in the same conditions
mentioned above.

Persistence is observed in the CorMASS detector when very bright
objects, e.g. ${\rm V} < 6$ calibration stars, are observed
spectroscopically.  ${\rm V} \geq 8$ calibrators are preferred for
this reason.

\section{Data Reduction \& Spectra}
\subsection{Data Reduction}
CorMASS first light observations occurred during 1999 August
22-25, with seven additional runs through 2000 August.

Single frame integration times of 100-300 seconds are typically
taken with telescope nods of $\sim 4.5 - 7\arcsec$ along the slit
between each integration. Frame integration times are reduced when
airglow and/or atmospheric water content are judged to be highly
variable to maintain effective sky subtraction.

Dome flats are taken before the start of each night's observing.
Dome illumination for the flats is provided by a 10W QTH (Quartz,
Tungsten and Halogen) lamp mounted above the telescope secondary.

Spectral observations of calibrator stars for atmospheric
correction are taken at airmasses within $\pm 0.1$ airmass of the
target observations. A-stars from the Elias standards
\citep{eli82} and bright late-F and G-stars have been used as
calibrators.

Following bad-pixel correction, the flat field and flux
calibration spectral images are corrected for NICMOS\,3 $shading$.
This is a noiseless exponential change in pixel bias based upon
position, time since last pixel read, and detector temperature.
The changing temperature of the detector readout amplifiers during
readout causes an exponentially varying bias in the direction
orthogonal to the pixel clocking, or along the columns
\citep{bok00}.  The columns are $\sim$ perpendicular to the orders
(Figure \ref{fpaimages}). To remove this artifact a quadratic is
fit to a row-by-row clipped median of the top quarter of the
array, the portion unused by the spectrograph. The fit is
extrapolated to 128 rows for the top two quadrants, duplicated for
the bottom two quadrants, and subtracted row-by-row from all
pixels.  The quadratic fit agrees well with row medians of
'quasi-dark' images taken with the flip mirror half-way between
spectrograph and slit-viewing mode.

Spectra are reduced using standard routines in IRAF\footnote{IRAF
is distributed by the National Optical Astronomy Observatories,
which are operated by the Association of Universities for Research
in Astronomy, Inc., under cooperative agreement with the National
Science Foundation.}.  Images are flat fielded using a pixel
responsivity solution derived from the APFLATTEN task using dome
flats from all nights of the run summed together.  Nodded image
pairs for science objects are subtracted against each other to
remove sky background to first-order and shading effects.  The
shading correction discussed above is not applied to science
object images because the subtraction of adjacent nodded images
removes the effect to a comparable noise level as the quadratic
fit method.

The APALL task is then used for spectral extraction. Wavelength
calibration is accomplished by using line identifications from
spectral observations of planetary nebulae, e.g. NGC 7027 and NGC
6210.  A third-order polynomial is fit to the identified lines to
derive a dispersion solution across the orders.  The fit residual
RMS is $\leq \case{1}{5}$ pixel ($2.25\AA$) across the wavelength
coverage. The use of OH airglow lines was rejected because at the
resolution of CorMASS ($R\sim300$), many airglow lines are blended
together, especially in the H-band, making line identifications
problematic.

Telluric absorption effects are removed by dividing by the reduced
spectrum of a standard observed at similar airmass and hand
corrected for stellar H Paschen and Brackett recombination lines
as required. The ratio is then multiplied by a black-body with a
temperature equal to the effective temperature of the stellar type
of the calibrator observed \citep{tok00}.  Lastly, all
observations for each order are weighted by the spectra mean and
combined with the SCOMBINE task using a trimmed average. Orders
are stitched together with SCOMBINE after hand deletion of noisy
data at the ends of the orders.

\subsection{Spectra}
\begin{figure}
\figurenum{11}
\label{ngc7027}
\plotone{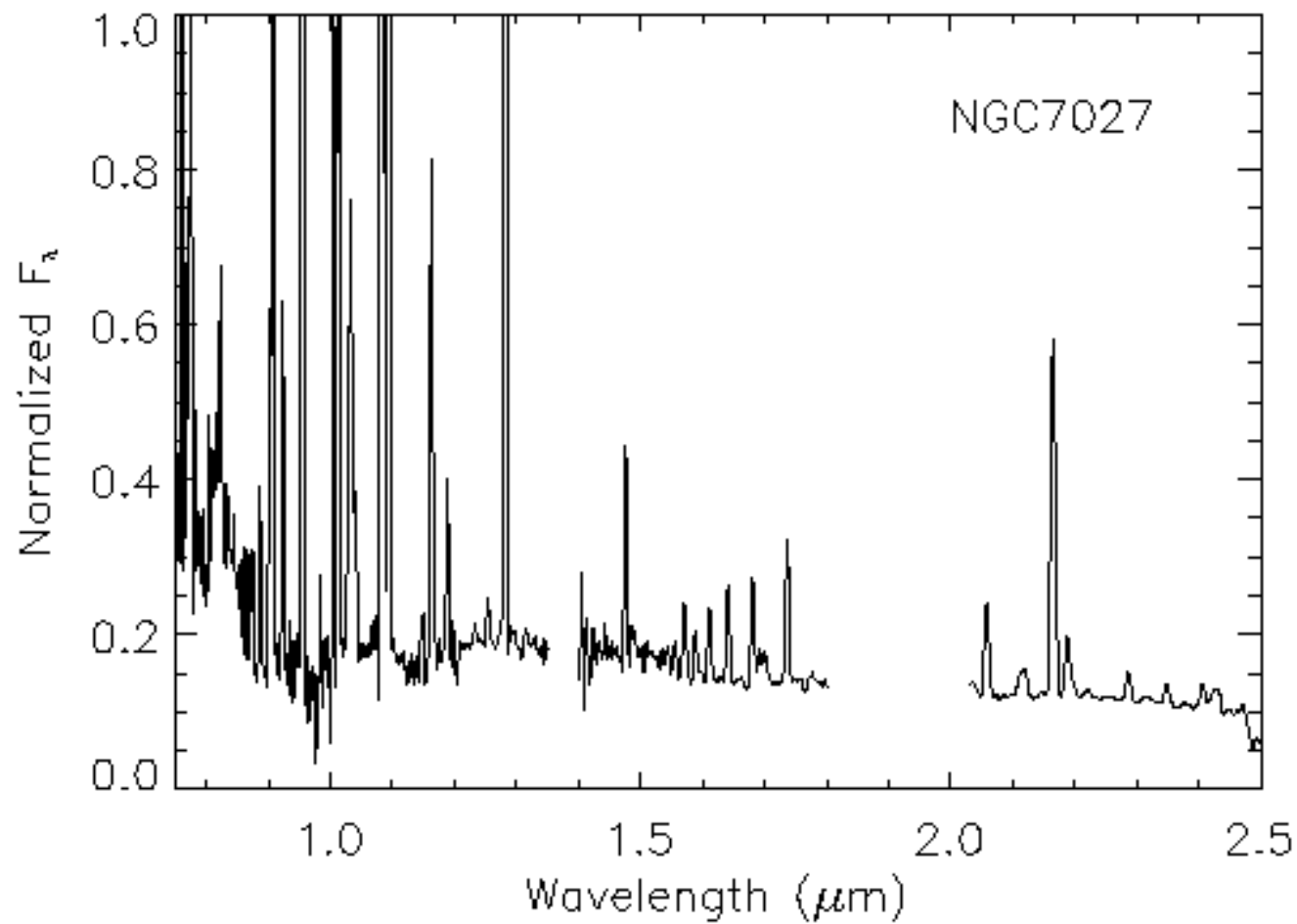}
\caption{CorMASS spectra of NGC 7027. This well
studied planetary nebula was observed 24 October 99 for ${\rm
t_{int}}=20$ sec.  38 distinct emission lines across 6 orders were
identified for use within IRAF to derive a dispersion solution for
the wavelength calibration of program objects. This plot has been
clipped to enhance fainter lines.}
\end{figure}

A spectrum of the infrared bright planetary nebula NGC 7027 is
shown in Figure \ref{ngc7027}. The plentiful emission lines of
planetary nebulae, primarily H and He lines, are used for
wavelength calibration.  CorMASS' wavelength coverage and
efficiency are well suited for the study of numerous types of
objects including quasars, novae, and emission line objects.

\begin{figure}
\figurenum{12}
\label{sequence}
\plotone{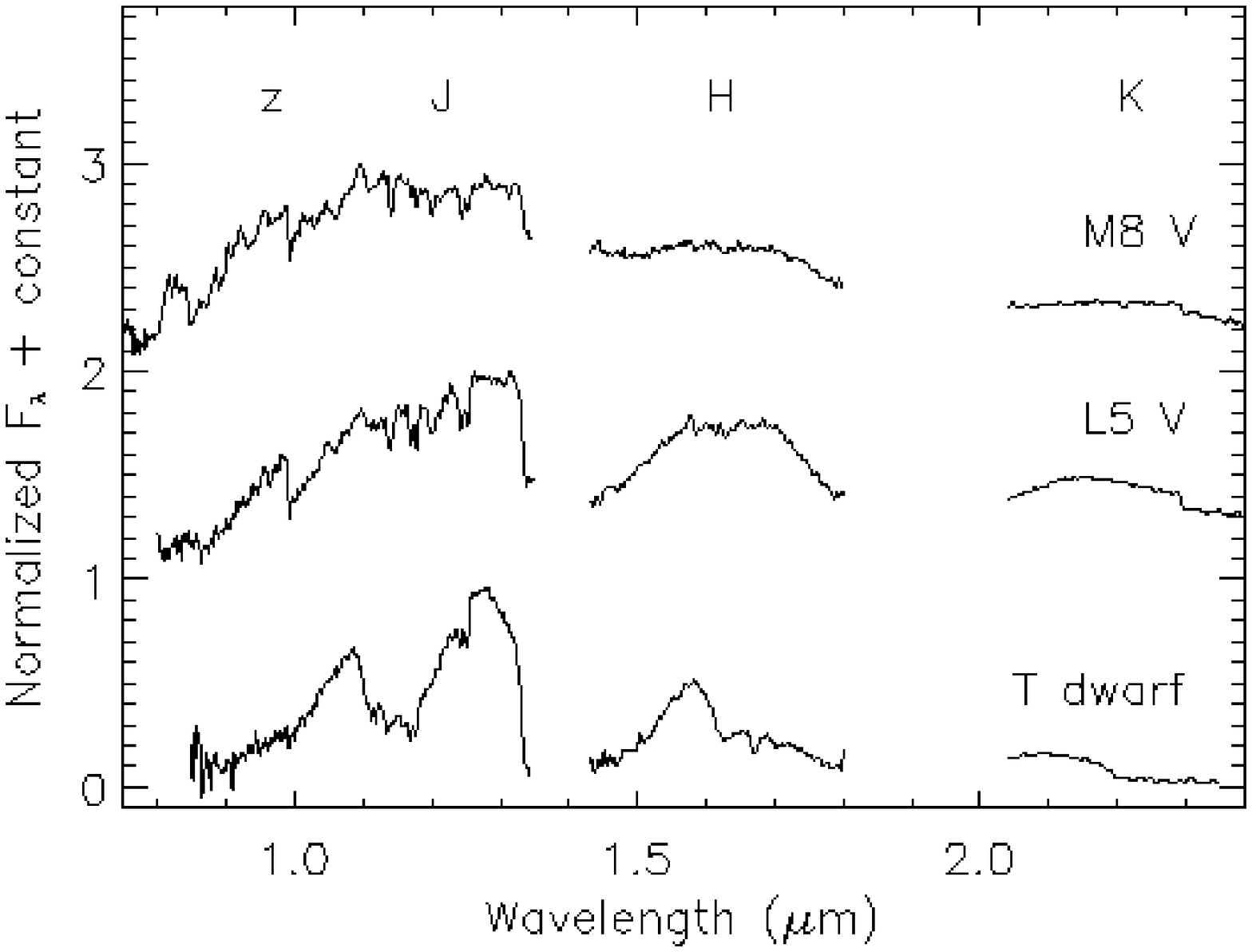}
\caption{CorMASS \textit{\'{z}JHK} spectra of three
low-mass objects: M 8V (VB 10), L 5V (2MASSW J1507476-162738,
hereafter J1507), and a T-dwarf (2MASSI J0559191-140448, hereafter
J0559). VB 10 ($K=8.80$, \citet{leg92}), was observed 22 August 99
for ${\rm t_{int}}=1080$ sec. J1507 ($K_s=11.30$, \citet{ki00a}),
was observed 20 April 00 for ${\rm t_{int}}=1800$ sec.  J0559
($K_s=13.61$, \citet{bur00}), was observed 24 October 99 for ${\rm
t_{int}}=2400$ sec.  Data between bands was not usable due to poor
atmospheric subtraction.}
\end{figure}

Spectra of the low-mass objects VB 10 (M 8V), 2MASSW
J1507476-162738 (L 5V, hereafter 2M1507), and the T-dwarf 2MASSI
J0559191-140448 (hereafter 2M0559) are presented in Figure
\ref{sequence}. A full spectral sequence and NIR L-dwarf features
will be discussed in another paper \citep{wil00}.

CorMASS spectroscopically confirmed the classification of 2M0559
($J=13.83\pm0.03$) during the instrument's second run
\citep{bur00}. 2M0559 is 0.4 mag brighter than Gl 229B and more
than 1 mag brighter than the other field T-dwarfs discovered to
date. This object's brightness will aid broad wavelength and high
resolution study of its spectral class.

While no methane combination and overtone absorption features at
1.6 and $2.2\micron$ (the features that define the T-dwarf class)
are evident in 2M1507, weak fundamental methane absorption has
been discovered at $3.3\micron$ in 2M1507 and in an L 7.5
\citep{nol00}.  This suggests that the transition from CO to ${\rm
CH}_4$ as the primary carbon-bearing molecule as effective
temperatures decrease commences as early as L 5V.

\section{Conclusion}
This paper described the newly built and commissioned NIR
low-resolution ($R\sim300$) prism cross-dispersed spectrograph
CorMASS in operation on the Palomar 60-inch telescope. This
instrument was designed primarily for the identification of late-M
and L-dwarf spectral features.  CorMASS spectra of low-mass
objects from the L, M, and T-dwarf classes, as well as the
planetary nebula NGC 7027, were presented. Instrument
characteristics include:\\ \indent1. Low-cost compact design.\\
\indent2. Simultaneous $\lambda$ coverage of the \textit{\'{z}JHK}
NIR bands with R $\sim 300$.\\ \indent3. A two-position flip
mirror for imaging an alternate slit viewing train.\\ \indent4.
Spectrograph sensitivity of $K=14.0$ in $t_{int}=3600$ sec for
S/N=5.
\\

\acknowledgements We wish to thank the Cornell University LASSP
(Laboratory of Applied and Solid-State Physics) Machine Shop and
the UMASS Physics Machine Shop for their professional and timely
machining. Many thanks to Hal Petrie for providing drawings and
important information about the 60-inch along the way. We also
wish to acknowledge Bob Thicksten and the rest of the Palomar
Observatory Staff for their wonderful attitudes and invaluable
assistance with the interface of CorMASS to the telescope and
professional telescope operations. Special thanks to Night
Assistants Skip Staples and Karl Dunscombe for their expertise and
support as we learned how to observe with this new instrument. It
is truly an enjoyable experience observing on Palomar Mountain. We
also wish to thank Eric Persson and David Murphy for information
on the Palomar 60-inch prescription and instrument mounting
constraints.  We largely duplicated the mounting scheme of their
P60 IR camera. Thanks to Tom Murphy for making available his list
of late-F and G-star calibration standards. We benefited from
useful discussions with Scott Milligan of Telic Optics on optical
design and element properties, as well as the thorough final
report for the 2MASS Prototype Camera redesign produced by his
company. We greatly benefited from Bernhard Brandl's advice and
experience with optical and mechanical design, as well as his
careful reading of a draft of this paper. We also wish to
acknowledge the influence of Jeff Van Cleve and the SCORE (SIRTF
Cornell Echelle Spectrograph) philosophy of a cross-dispersed
echelle with minimal moving parts. JCW and JDS acknowledge support
by NASA grant NAG5-4376. This publication makes use of data from
the Two Micron All Sky Survey, which is a joint project of the
University of Massachusetts and the Infrared Processing and
Analysis Center, funded by the National Aeronautics and Space
Administration and the National Science Foundation.

\clearpage

\clearpage

\begin{deluxetable}{lcccc}
\tabletypesize{\scriptsize} \tablewidth{0pt} \tablecaption{Palomar
60-inch and Spectrograph Prescription \label{tbl-sp}}
\tablehead{\colhead{Surface} & \colhead{Radius\tablenotemark{a} \
(mm)} & \colhead{Thickness\tablenotemark{a,b} \ \ \ (mm)} &
\colhead{Material} & \colhead{Diameter (mm)}}

\startdata Central Obscuration & \nodata & $\sim 2900$ & \nodata &
    640.08\\
Tel Primary\tablenotemark{c} & -7619.977 &
    -2827.918 & coated Alum & 1549.40\\
Tel Secondary\tablenotemark{d} & -2750.003 & 3402.212 & coated
    Alum & 454.56\\
Cass Mtg Surf & \nodata & 3.917 & \nodata & \nodata\\ Dewar Window
(R1) & Inf & 3.074 & AR coated ${\rm CaF}_2$ & 25.40\\ Dewar
Window (R2) & Inf & 28.444 & \nodata & 25.40\\ Reflective
Slit\tablenotemark{e} & Inf & 114.277 & SS 304 & 9.53\\
f/Converter Lens 1 (R1) & -166.870 & 3.000 & AR coated Infrasil
301 & 38.10\\ f/Converter Lens 1 (R2) & Inf & 4.189 & \nodata &
38.10\\ f/Converter Lens 2 (R1) & Inf & 4.502 & AR coated ${\rm
CaF}_2$ & 38.10\\ f/Converter Lens 2 (R2) & -79.810 & 101.600 &
\nodata & 38.10\\ Off-Axis Paraboloid\tablenotemark{f} & -60.960 &
-143.479 & Au coated Alum
    & 38.10\\
Lyot Stop & \nodata & -46.521 & \nodata & 22.56\\ Prism 1st Pass
(R1)\tablenotemark{g} & Inf & -20.000 & SF-10 & \nodata\\ Prism
1st Pass (R2) & Inf & -30.000 & \nodata & \nodata\\
Grating\tablenotemark{h} & Inf & 30.000 & Au coated Alum &
\nodata\\ Prism 2nd Pass (R2) & Inf & 20.000 & SF-10 & \nodata\\
Prism 2nd Pass (R1) & Inf & 75.000 & \nodata & \nodata\\ Camera
Lens 1 (R1) & 109.520 & 8.647 & AR coated ${\rm BaF}_2$ & 50.80\\
Camera Lens 1 (R2) & -120.289 & 4.460 & \nodata & 50.80\\ Camera
Lens 2 (R1) & -94.600 & 10.000 & AR coated Infrasil 301 & 50.80\\
Camera Lens 2 (R2) & -287.900 & 131.430 & \nodata & 50.80\\ Camera
Lens 3 (R1)\tablenotemark{i} & 36.540 & 10.000 & AR coated
Infrasil 301 & 50.80\\ Camera Lens 3 (R2) & 258.750 & 23.569 &
\nodata & 50.80\\ Detector & \nodata & \nodata & NICMOS\,3 &
$10.24^2$
\enddata

\tablenotetext{a}{Cold Dimensions (77K) for slit and below}
\tablenotetext{b}{Distance to next face or element}
\tablenotetext{c}{Conic Constant: $k=-1$; Sag (mm):
    $(-6.5617e-5)y^2 + (1.835e-14)y^4$}
\tablenotetext{d}{Conic Constant: $k=-1$; Sag (mm): $(-1.81818e-4)y^2 + (1.743e-11)y^4$}
\tablenotetext{e}{Slit rotated $45 \deg$ to optical axis along
    slit axis}
\tablenotetext{f}{Focal Length (cold) $=12 \arcsec$,
    Decenter (cold) $=115.57 {\rm mm}$, Off-Axis Diameter (OAD) (cold) $=3.8 \arcsec$}
\tablenotetext{g}{$30 \deg$ Prism apex angle, used in minimum
    deviation} \tablenotetext{h}{$40 {\rm lines}/{\rm mm}$,
    $\lambda_{blaze}=4.8\micron$, Blaze Angle $=5.5\deg$, Richardson Grating Lab replica}
\tablenotetext{i}{Camera Lens 3 is shared with the Slit Viewing
    path}
\end{deluxetable}

\begin{deluxetable}{lcccc}
\tabletypesize{\scriptsize} \tablewidth{0pt} \tablecaption{Slit
Viewer Prescription \label{tbl-sv}} \tablehead{\colhead{Surface} &
\colhead{Radius\tablenotemark{a} \ (mm)} &
\colhead{Thickness\tablenotemark{a,b} \ \ \ (mm)} &
\colhead{Material} & \colhead{Diameter (mm)}}

\startdata Reflective Slit\tablenotemark{c} & Inf & -41.287 & SS
304 & 9.53\\ Slit Vwr Lens 1 (R1) & Inf & -4.558 & ${\rm CaF}_2$ &
19.05\\ Slit Vwr Lens 1 (R2) & 19.018 & -45.146 & \nodata &
19.05\\ Lyot Stop & Inf & -46.126 & \nodata & 5.111\\ Fold Mirror
1 & Inf & 50.8 & Au coated BK7 & 19.05\\ $K_s$ Filter
(R1)\tablenotemark{d} & Inf & 2.0 & \nodata & 25.4\\ $K_s$ Filter
(R2) & Inf & 193.146 & \nodata & 25.4\\ Fold Mirror 2 & Inf &
-18.170 & Au coated Pyrex & 31.42\\ Slit Vwr Lens 2 (R1) & -59.819
& -3.390 & ${\rm CaF}_2$ & 25.4\\ Slit Vwr Lens 2 (R2) & Inf &
-35.118 & \nodata & 25.4\\ Flip Mirror & Inf & 55.494 & Au coated
Pyrex & 25.4\\ Camera Lens 3 (R1)\tablenotemark{e} & 36.540 &
10.000 & AR coated Infrasil 301 & 50.80\\ Camera Lens 3 (R2) &
258.750 & 23.569 & \nodata & 50.80\\ Detector & \nodata & \nodata
& NICMOS\,3 & $10.24^2$
\enddata

\tablenotetext{a}{Cold Dimensions (77K) for slit and below}
\tablenotetext{b}{Distance to next face or element}
\tablenotetext{c}{Slit rotated $45 \deg$ to optical axis along
slit axis} \tablenotetext{d}{Filter tilted $5 \deg$ towards side
of Dewar for stray light control} \tablenotetext{e}{Camera Lens 3
is shared with the Spectrograph path}
\end{deluxetable}

\begin{deluxetable}{lcc}
\tabletypesize{\small} \tablewidth{0pt} \tablecaption{Coefficients
of Thermal Expansion \label{tbl-cle}}
\tablehead{\colhead{Material} & \colhead{$\Delta L/L
(\%)$\tablenotemark{a}} & \colhead{Reference}}

\startdata Alum & 0.392 & \citet{cas93} \& \citet{tou75}\\ ${\rm
CaF}_2$ & 0.303 & \citet{cas93} \& \citet{tou75}\\ ${\rm BaF}_2$ &
0.319 & \citet{cas93} \& \citet{tou75}\\ Infrasil-301 & 0.0 & Type
I ${\rm SiO}_2$ (Fused), \citet{tou75}\\ SF-10 & 0.127 & Assume
$(\Delta L/L)_{SF-10} \sim (\Delta L/L)_{Al} * \alpha(0\degr
C)_{SF-10}/\alpha(0\degr C)_{Al}$
\enddata
\tablenotetext{a}{293K-77K}

\end{deluxetable}

\begin{deluxetable}{lc}
\tabletypesize{\small} \tablewidth{0pt} \tablecaption{CorMASS
Performance \label{performance}} \tablehead{\colhead{Parameter} &
\colhead{Value}}

\startdata  Detector & NICMOS\,3 \\ Bias & 0.91 V
\\ Gain & 8.5 $e^{-}/{\rm DN}$ \\ Read Noise & 42.5 $e^{-}$ \\
Well Depth & $\sim 3.8$ x $10^5$ $e^{-}$ \\ Bad Pixels & $< 1 \%$
\\ Spec Sensitivity\tablenotemark{a,b} & $J=14.6$, $H=14.9$, $K=14.0$
\\ Spec Ave Throughput\tablenotemark{b,c}
& $\textit{\'{z}}=0.05$, $J=0.07$, $H=0.13$, $K=0.10$
\\ Slit Viewer Sensitivity\tablenotemark{b} & $K_s=13$ in 2 sec
for $\sigma = 5$ \\ Slit Viewer Throughput\tablenotemark{c} & 0.20
\enddata
\tablenotetext{a}{S/N=5, $t_{int}=3600$ sec} \tablenotetext{b}{For
a point source detection with a subtracted background in clear
conditions with seeing $\la2.0\arcsec$ at K.}
\tablenotetext{c}{$e^{-}/{\rm photon}$, includes transmission
losses from the atmosphere, telescope, instrument and detector
QE.}
\end{deluxetable}

\end{document}